# The three-dimensional statistical characterization of plain grinding surfaces

LIANG Xuan-ming, YUAN Wei-ke, DING Yue, WANG Gang-feng[✉]

(Department of Engineering Mechanics, SVL, Xi'an Jiaotong University, Xi'an, 710049, China)

**Abstract:** In tribology, it is of importance to properly characterize the topography of rough surfaces. In this work, the three-dimensional topographies of plain grinding surfaces are measured through a white light interferometer, and their geometrical statistical features are analyzed. It is noticed that only when the total measured area is larger than a threshold value, is the statistical characterization reasonable and stable, which should be kept in mind in actual measurements. For various plain grinding surfaces, the height of asperity-summit obeys a Gaussian distribution, and the equivalent curvature radius follows a modified *F*-distribution. These statistical characteristics are helpful to analyze the contact and friction behaviors of rough surfaces.

## 1 Introduction

Surface topography has a significant influence on the mechanical and physical performances of contacting bodies, such as friction, wear, lubrication, sealing, electricity transfer, heat transfer, etc.[1,2] Therefore, much effort has been devoted to characterizing the topography of rough surfaces in the past decades. However, it is still a challenging task on account of the geometric complexity and randomness.

Among all those characterizing methods, the multi-asperity models and the fractal models have attracted most attention. In multi-asperity models, if the height and size of each asperity over the rough surfaces are known, then the overall contact response can be achieved by simply summing over that of each single asperity. Early in 1957, Archard[3] modeled the rough surface as an ideal hierarchical structure with smaller spherical asperities on larger spherical asperities and revealed the power-law relationship between the real contact area and the external load. For bead-blasted aluminum surfaces, Greenwood and Williamson[4] found that the height of asperity-summit follows a Gaussian distribution. Then assuming an identical summit curvature radius, they established the famous theoretical rough surface contact model (GW model). In 1958, Longuet-Higgins[5] derived the statistical distribution of the curvature of random Gaussian surfaces from their power spectrum density. This model was further developed by Nayak[6] to account for various statistical features of rough surfaces, such as the height and curvature of summit. Later, Bush et al.[7] approximated the summits by elliptic paraboloids and employed Nayak's random process theory to characterize the distributions of height and principal curvature.

Corresponding author. wanggf@mail.xjtu.edu.cn

In 1984, Mandelbrot et al.[8] observed that the roughness of fractured surfaces appears quite similar under various degrees of magnification, and revealed the fractal nature of rough surfaces. Majumdar and Tien[9] therefore introduced the Weierstrass-Mandelbrot function to characterize this kind of self-affine rough surfaces. Thereafter, many researches have been conducted on the fractal contact models[10-12].

The aforementioned description methods mainly focus on the isotropic rough surfaces, but ignore the anisotropy trait which is always unavoidable for many practical machined metal surfaces. Theoretically, the statistical geometry based on random process theory can completely describe the general anisotropic Gaussian rough surfaces with a great number of parameters involved. With some practical assumptions introduced, this complicated theory has been slightly simplified. Bush et al.[13] suggested that the asperities on the strongly anisotropic surfaces can be represented by highly eccentric paraboloids with their semi-major axes oriented in the same direction. So and Liu[14] proposed a more general description method by only assuming the profiles of asperities along machined direction have minimum slope and curvature. Nonetheless, the contact solutions of these works are still very cumbersome, and the calculations of the real contact area are difficult to implement. If further simplifications referring to the idea of Greenwood[4] can be carried out for the anisotropic rough surfaces, such complexity might be avoided with a small accuracy cost.

In addition to the theoretical models mainly based on assumptions and conjectures, experimental approaches aiming at the accurate description of rough surfaces have also been conducted. With the profiles measured from engineering surfaces, Aramaki et al.[15] and Ciulli et al.[16] employed a reference line to truncate the measured profile into discrete ones and defined each truncated isolated part above the reference line as an asperity. This identification criterion of asperity suits well for 2D profile, but cannot be simply extended to 3D topography. Recently, Kalin and Pogačnik[17] measured the 3D topographies of steel specimens, and discussed the influence of asperity identification criteria on the characterization of rough surfaces. However, few analyses are aiming at the statistical distributions of rough surfaces based on these experimental results.

In this study, the experimental measurement of 3D topographies combined with statistical analysis is performed for the characterization of plain grinding rough surfaces. A new asperity-summit identification criterion is adopted to determine the 3D geometrical properties of asperities. Based on sufficient valid surface data, the probability distributions of height and equivalent radius of asperities are generalized, which are helpful to develop simplified contact models for anisotropic rough surface.

## 2 Experimental method

Surface specimens were prepared by applying standard plain grinding processing on the 45 quality carbon structural steel (C45E4, ISO 683-18-1996). Six specimens with increasing roughness were produced as shown in Fig. 1, and their size is 20 mm × 8 mm. In this work, the 3D surface topography was recorded by using a white light interferometer (NanoMap-1000WLI, AEP). A 50× interference objective lens and a 1024 × 1024 pixel CCD sensor were used. Thus surface topography of a square region 209.6 × 209.6 μm² was measured in each scanning procedure with a lateral resolution of 0.2 μm.

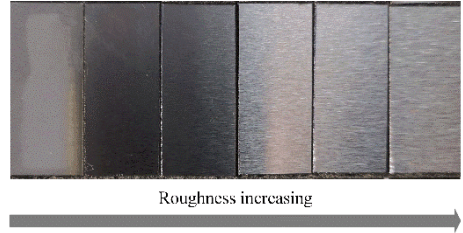

Fig.1　Surface specimens prepared by standard plain grinding processing

Fig. 2 shows a typical scanning result of the plain grinding surfaces. Clear distinction is observed between *x*-direction and *y*-direction, which indicates the anisotropy of such kind of rough surfaces. It should be pointed out that the scanned region in each operation is much smaller than the specimen size. To obtain the statistical properties of a whole specimen surface, more than 50 such squared regions were measured for each specimen, which means the total scanned area is over 2 mm². To ensure the statistical and random nature, a randomized algorithm was utilized to choose the positions for scanning. Note that the surface altitude on the same specimen was recorded with an identical reference plane, for the convenience to summarize the distribution of summit height in a global consideration.

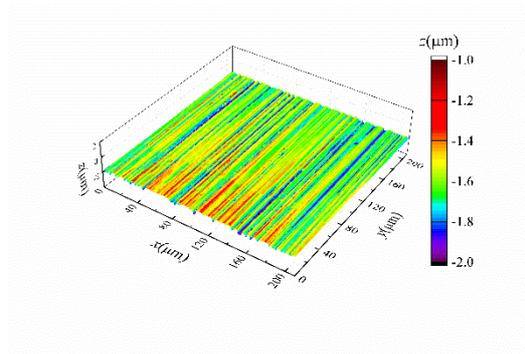

Fig.2　A typical scanning result of plain grinding surfaces with a 50× interference objective lens

By applying the above approach, six specimens with different surface roughness were measured separately. For each specimen, the height of the *n*-th scanning position is denoted as $z_n(i, j)$, where $i$ and $j$ are the discrete coordinates of sampling points varying from 1 to 1024. The reference plane was chosen to make sure the mean value of height equals zero. Thus the surface roughness can be calculated by

$$S_a = \frac{1}{1024^2 N} \sum_{n=1}^{N} \sum_{i=1}^{1024} \sum_{j=1}^{1024} |z_n(i,j)| \qquad (1)$$

where $N$ denotes the total number of scanned regions. Our analysis gave the surface roughness $S_a$ of six specimens as 188 nm, 432 nm, 610 nm, 720 nm, 823 nm, and 900 nm, respectively.

According to our scanning result, the peak of one asperity may consist of several points instead of one single point as shown in Fig. 3. Table 1 shows the measured height data of a typical asperity-summit. Note that, for a single asperity-summit, there are 9 peak points having the same height value −1496 nm, which is higher than those of surrounding points. For a summit with only one highest point, the 9-point rectangular approach[17] was widely used to calculate its curvature. However, for asperity-summits with a cluster of peak points, the 9-point rectangular approach is no longer applicable, and it is necessary to develop a new criterion to identify each asperity-summit and calculate its height and principal curvature. In this work, we use an elliptic paraboloid to approximate each asperity-summit consisting of multiple peak points (the red points in Fig. 3) and neighboring points (the yellow points in Fig. 3). In the local cylindrical coordinate system of which the origin locates at the projection center of the asperity-summit onto the reference plane, the elliptic paraboloid $z(\rho, \theta)$ can be expressed as

$$z = h - \frac{\rho^2}{4}\{k_1 + k_2 + (k_1 - k_2)\cos[2(\alpha - \theta)]\} \qquad (2)$$

where $k_1$ and $k_2$ represent the bigger and the smaller principal curvature, respectively, $\alpha$ indicates the principal direction, and $h$ is the height of the asperity-summit. Through a fitting procedure, we can obtain the height and principal curvature of each asperity-summit.

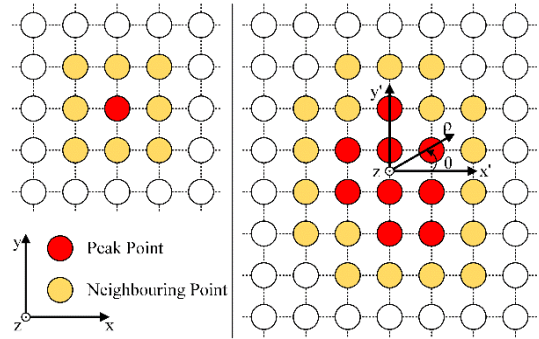

Fig.3 Identification of the asperity summits on 3D surfaces: left for 9-point rectangular asperity summit[17], right for an irregular asperity summit

Table 1  Height data $z$ of a typical asperity summit (nm)

| j \ i | 35 | 36 | 37 | 38 | 39 | 40 | 41 |
|---|---|---|---|---|---|---|---|
| 34 | −1624 | −1537 | −1514 | −1514 | −1581 | −1623 | −1710 |
| 35 | −1624 | −1503 | −1500 | −1500 | −1500 | −1581 | −1710 |
| 36 | −1624 | −1500 | −1498 | −1496 | −1498 | −1581 | −1710 |
| 37 | −1649 | −1500 | −1496 | −1496 | −1496 | −1503 | −1705 |
| 38 | −1649 | −1511 | −1496 | −1496 | −1496 | −1503 | −1663 |
| 39 | −1649 | −1531 | −1503 | −1496 | −1496 | −1503 | −1656 |
| 40 | −1649 | −1531 | −1503 | −1503 | −1503 | −1511 | −1582 |
| 41 | −1625 | −1538 | −1511 | −1503 | −1503 | −1511 | −1579 |

## 3  Results and discussion

Before summarizing the statistic information of rough surface, a necessary preparation is conducted to check how large a scanned area is required, which is always neglected in previous works. Fig. 4 displays the dependence of the standard deviation of summit height $\sigma_h$ on the total scanning area $A$. When $A$ is small, $\sigma_h$ varies essentially with the scanned position. Such fluctuations are represented by the error bars. However, as the total scanned area $A$ increases, $\sigma_h$ converges to a steady value independent of the scanned position. Moreover, the specimen with higher surface roughness requires a larger area $A$ to achieve a stable $\sigma_h$. For example, $\sigma_h$ of the specimen with $S_a$ = 188 nm stabilizes when $A$ is about 0.5 mm$^2$, while $\sigma_h$ of the specimen with $S_a$ = 900 nm becomes stable until $A$ increases to 2 mm$^2$. Other statistics like surface roughness and principal curvature exhibit similar characteristics. These results suggest that only when the scanned area in the surface topography measurement is sufficiently large, can the statistical information of the entire surface be collected reliably. Neglecting this effect and analyzing on a small scanned area may lead to irregular experimental results.

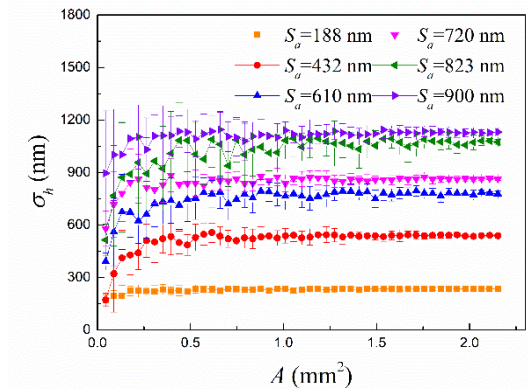

Fig.4  Variation of the standard deviation of summit height $\sigma_h$ with respect to the sampling area $A$

In this work, the statistical analyses of the geometric parameters of asperity-summits on each specimen surface are based on a sufficiently large area (2.15 mm$^2$ consisting of 50 randomly chosen scanning regions). The asperity-summit identification criterion and fitting method described in previous section were employed. By counting the height and curvature of all asperity-summits identified, we obtained the statistical distributions of various surface parameters.

Fig. 5 displays the distribution of summit height $h$. It is found that, for six specimens with different roughness, $h$ follows a Gaussian distribution, and can be expressed by

$$f(h) = \frac{1}{\sqrt{2\pi}\sigma} \exp\left(-\frac{(h-\mu)^2}{2\sigma^2}\right) \qquad (3)$$

where $\sigma$ and $\mu$ represent the standard deviation and the expectation of summit height, respectively. $\sigma$ and $\mu$ for six specimens are given in Table 2. Moreover, it is interesting to find that the standard deviation $\sigma$ exhibits linear dependence on $S_a$ and can be fitted by

$$\sigma/S_a = 1.2087 \qquad (4)$$

The absolute value of $\mu$ is equal to the distance between the summit mean plane and the reference plane. Since in our experiments, different reference planes are chosen for various specimens, it is not surprising to see various values of $\mu$. However, the exact value of $\mu$ has no impact on the load-area relationship in multi-asperity contact models.

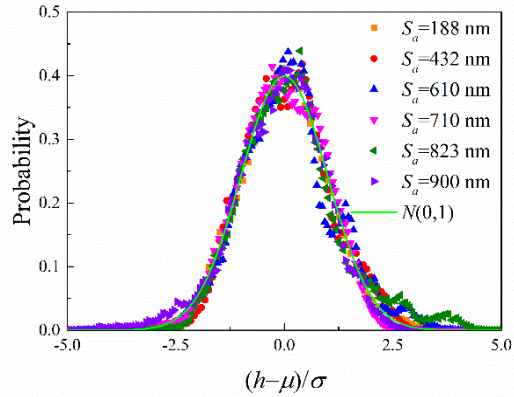

Fig.5 Probability density distribution of normalized summit height $(h-\mu)/\sigma$

Table 2  Fitting parameters of the probability density distributions of $h$ and $R_g$ (nm)

| $S_a$ | 188 | 432 | 610 | 720 | 823 | 900 |
|---|---|---|---|---|---|---|
| $\sigma$ | 227 | 542 | 727 | 921 | 937 | 1065 |
| $\mu$ | 72 | 37 | 75 | 187 | −10 | 285 |
| $p$ | 1632 | 4663 | 3204 | 2801 | 2871 | 2472 |

In addition to the summit height, the distribution of summit curvature radius is also analyzed. It is found that most ratios of the principal curvature $k_1/k_2$ are smaller than 10. For an asperity-summit like this, its elastic deformation can be approximately described by the simple spherical Hertzian solution with a small accuracy cost[18]. An available option is to replace the sphere radius of Hertzian solution with an equivalent radius $R_g$, which can be expressed as

$$R_g = (k_1 k_2)^{-1/2} \tag{5}$$

Following a similar procedure as dealing with summit height, we could obtain the statistical distribution of the equivalent radius $R_g$ as shown in Fig. 6. The probability density distribution of $R_g$ for each specimen can be fitted by a modified $F$-distribution as

$$g(R_g) = \begin{cases} \dfrac{\Gamma(\dfrac{n_1+n_2}{2})}{\Gamma(\dfrac{n_1}{2})\Gamma(\dfrac{n_2}{2})} (\dfrac{n_1}{n_2})(\dfrac{n_1}{n_2}\dfrac{R_g}{p})^{\frac{n_1}{2}-1}(1+\dfrac{n_1}{n_2}\dfrac{R_g}{p})^{-\frac{n_1+n_2}{2}}\dfrac{1}{p}, & R_g > 0 \\ 0 & , R_g \leq 0 \end{cases} \tag{6}$$

where $\Gamma$ is the Gamma function, $p$ is a scaling factor of $R_g$, $n_1$ and $n_2$ take values of 8.0 and 2.2, respectively. From this expression, we can conclude that the normalized equivalent radius $R_g/p$ follows an $F$-distribution with degrees of freedom being (8.0, 2.2). The values of $p$ for six specimens are given in Table 2.

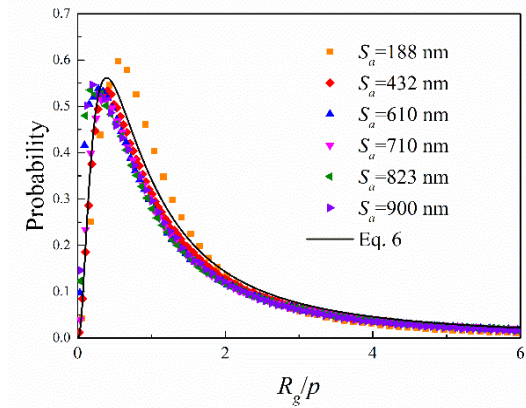

Fig.6　Probability density distribution of the normalized equivalent radius $R_g/p$

Moreover, from the data collected on each specimen, the joint probability density distribution of $h$ and $R_g$, denoted as $p(h, R_g)$, can be obtained numerically. Directly fitting $p(h, R_g)$ is rather difficult. If we further assume $h$ and $R_g$ are two independent random variables, we could calculate the joint probability density distribution by simply multiplying $f(h)$ with $g(R_g)$. For instance, the distributions $f(h) \times g(R_g)$ and $p(h, R_g)$ of the specimen with $S_a = 188$ nm are shown in Fig. 7a and 7b, respectively. It is seen that the distribution $f(h) \times g(R_g)$ has little distinction from $p(h, R_g)$. Similar results have been observed for other specimens. This comparison leads us to the conclusion that the statistical distributions of $h$ and $R_g$ are almost independent and their joint probability density distribution can be approximately expressed as

$$p(h, R_g) \approx f(h)g(R_g) \qquad (7)$$

The statistical distributions of summit height and curvature radius, and their joint distribution can be directly used to develop contact models for anisotropic plain grinding rough surfaces. Undoubtedly, this method can also be applied to conduct similar statistical analyses on other engineering surfaces.

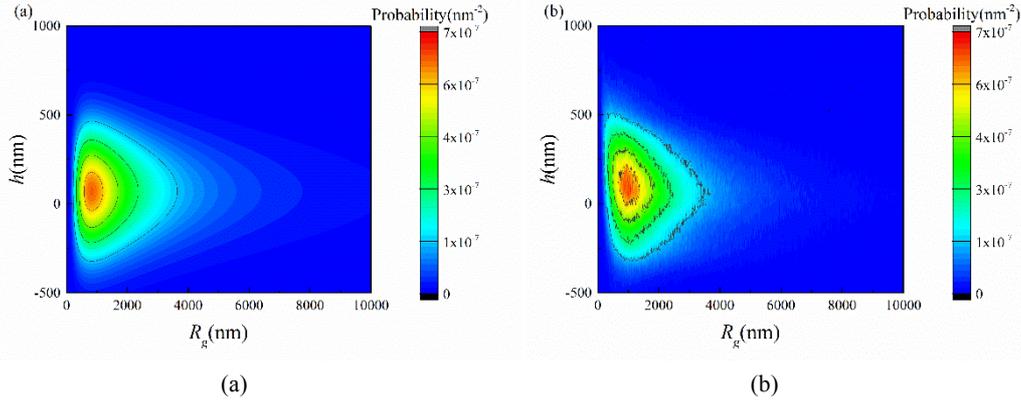

|  (a)  |  (b)  |

Fig.7 The joint probability density distribution of $h$ and $R_g$ (a) $f(h) \times g(R_g)$, (b) $p(h, R_g)$

## 4  Conclusion

The three-dimensional topographies of plain grinding surfaces are characterized by combing the experimental measurement with statistical analyses. The results suggest that the statistical characterization of rough surface should be conducted on a sufficiently large sampling area. With a new asperity identification criterion employed, it is found that the height of asperity-summit follows a Gaussian distribution, while the equivalent radius of asperity-summit follows a modified $F$-distribution. Moreover, we demonstrate the height and equivalent radius of asperity-summit are nearly independent, and the joint probability density function is derived directly. These findings can be utilized for developing simplified contact models of anisotropic rough surfaces and provide a new perspective for the characterization of rough surfaces.